\PassOptionsToPackage{unicode}{hyperref}
\PassOptionsToPackage{hyphens}{url}
\PassOptionsToPackage{dvipsnames,svgnames,x11names}{xcolor}
\documentclass[
]{article}

\usepackage{amsmath,amssymb}
\usepackage{iftex}
\ifPDFTeX
  \usepackage[T1]{fontenc}
  \usepackage[utf8]{inputenc}
  \usepackage{textcomp} 
\else 
  \usepackage{unicode-math}
  \defaultfontfeatures{Scale=MatchLowercase}
  \defaultfontfeatures[\rmfamily]{Ligatures=TeX,Scale=1}
\fi
\usepackage{lmodern}
\ifPDFTeX\else  
\fi
\IfFileExists{upquote.sty}{\usepackage{upquote}}{}
\IfFileExists{microtype.sty}{
  \usepackage[]{microtype}
  \UseMicrotypeSet[protrusion]{basicmath} 
}{}
\makeatletter
\@ifundefined{KOMAClassName}{
  \IfFileExists{parskip.sty}{%
    \usepackage{parskip}
  }{
    \setlength{\parindent}{0pt}
    \setlength{\parskip}{6pt plus 2pt minus 1pt}}
}{
  \KOMAoptions{parskip=half}}
\makeatother
\usepackage{xcolor}
\ifx\paragraph\undefined\else
  \let\oldparagraph\paragraph
  \renewcommand{\paragraph}[1]{\oldparagraph{#1}\mbox{}}
\fi
\ifx\subparagraph\undefined\else
  \let\oldsubparagraph\subparagraph
  \renewcommand{\subparagraph}[1]{\oldsubparagraph{#1}\mbox{}}
\fi

\usepackage{longtable,booktabs,array}
\usepackage{calc} 
\usepackage{etoolbox}
\makeatletter
\patchcmd\longtable{\par}{\if@noskipsec\mbox{}\fi\par}{}{}
\makeatother
\IfFileExists{footnotehyper.sty}{\usepackage{footnotehyper}}{\usepackage{footnote}}
\makesavenoteenv{longtable}
\usepackage{graphicx}
\makeatletter
\def\maxwidth{\ifdim\Gin@nat@width>\linewidth\linewidth\else\Gin@nat@width\fi}
\def\maxheight{\ifdim\Gin@nat@height>\textheight\textheight\else\Gin@nat@height\fi}
\makeatother
\setkeys{Gin}{width=\maxwidth,height=\maxheight,keepaspectratio}
\makeatletter
\def\fps@figure{htbp}
\makeatother
\NewDocumentCommand\citeproctext{}{}

\makeatletter
 \let\@cite@ofmt\@firstofone
 \def\@biblabel#1{}
 \def\@cite#1#2{{#1\if@tempswa , #2\fi}}
\makeatother
\newlength{\cslhangindent}
\newlength{\csllabelwidth}
\newenvironment{CSLReferences}[2] 
 {\begin{list}{}{%
  \setlength{\itemindent}{0pt}
  \setlength{\leftmargin}{0pt}
  \setlength{\parsep}{0pt}
  \ifodd #1
   \setlength{\leftmargin}{\cslhangindent}
   \setlength{\itemindent}{-1\cslhangindent}
  \fi
  \setlength{\itemsep}{#2\baselineskip}}}
 {\end{list}}
\usepackage{calc}

\usepackage{arxiv}
\usepackage{orcidlink}
\usepackage{amsmath}
\usepackage[T1]{fontenc}
\usepackage[utf8]{inputenc}
\usepackage{pifont}
\usepackage{newunicodechar}
\usepackage{amssymb}
\newunicodechar{✓}{\ding{51}}
\newunicodechar{✗}{\ding{55}}
\newunicodechar{̆}{\u{}}
\newunicodechar{∈}{\in}
\newcommand{\downarrowcorner}{\rotatebox[origin=c]{180}{$\Lsh$}}
\newunicodechar{↳}{\downarrowcorner}
\makeatletter
\@ifpackageloaded{caption}{}{\usepackage{caption}}
\AtBeginDocument{%
\ifdefined\contentsname
  \renewcommand*\contentsname{Table of contents}
\else
  \newcommand\contentsname{Table of contents}
\fi
\ifdefined\listfigurename
  \renewcommand*\listfigurename{List of Figures}
\else
  \newcommand\listfigurename{List of Figures}
\fi
\ifdefined\listtablename
  \renewcommand*\listtablename{List of Tables}
\else
  \newcommand\listtablename{List of Tables}
\fi
\ifdefined\figurename
  \renewcommand*\figurename{Figure}
\else
  \newcommand\figurename{Figure}
\fi
\ifdefined\tablename
  \renewcommand*\tablename{Table}
\else
  \newcommand\tablename{Table}
\fi
}
\@ifpackageloaded{float}{}{\usepackage{float}}
\floatstyle{ruled}
\@ifundefined{c@chapter}{\newfloat{codelisting}{h}{lop}}{\newfloat{codelisting}{h}{lop}[chapter]}
\floatname{codelisting}{Listing}

\makeatother
\makeatletter
\@ifpackageloaded{caption}{}{\usepackage{caption}}
\@ifpackageloaded{subcaption}{}{\usepackage{subcaption}}
\makeatother
\ifLuaTeX
  \usepackage{selnolig}  
\fi
\usepackage{bookmark}

\IfFileExists{xurl.sty}{\usepackage{xurl}}{} 
\hypersetup{
  pdftitle={ICBeLLM: High Quality International Events Data with Open Source Large Language Models on Consumer Hardware},
  colorlinks=true,
  linkcolor={blue},
  filecolor={Maroon},
  citecolor={Blue},
  urlcolor={Blue},
  pdfcreator={LaTeX via pandoc}}

\newcommand{\runninghead}{A Preprint }
\title{ICBeLLM: High Quality International Events Data with Open Source
Large Language Models on Consumer Hardware}
\def\asep{\\\\\\ } 
\def\asep{\And }
\author{\textbf{Rex W. Douglass}\\\\University of California, San
Diego\\\\\asep\textbf{Thomas Leo Scherer}\\\\University of California,
San Diego\\\\\asep\textbf{J. Andrés Gannon}\\\\Vanderbilt
University\\\\\asep\textbf{Erik Gartzke}\\\\University of California,
San Diego\\\\}
\date{}
\begin{document}
\maketitle
\begin{abstract}
The International Crises Behavior Events (ICBe) ontology provides high
coverage over the thoughts, communications, and actions that constitute
international relations. A major disadvantage of that level of detail is
that it requires large human capital costs to apply it manually to new
texts. Whether such an ontolgy is practical for international relations
research given limited human and financial resources is a pressing
concern. We introduce a working proof of concept showing that ICBe
codings can be reliably extracted from new texts using the current
generation of open source large language models (LLM) running on
consumer grade computer hardware. Our solution requires no finetuning
and only limited prompt engineering. We detail our solution and present
benchmarks against the original ICBe codings. We conclude by discussing
the implications of very high quality event coding of any text being
within reach of individual researchers with limited resources.
\end{abstract}

\section{Introduction}\label{introduction}

The International Crisis Behavior Events (ICBe) project provides a
sentence-event level measurement of most thoughts, speech, and actions
described by a corpus of historical narratives of an international
crises (Douglass et al. 2022). The ontology represents the current state
of the art in human extraction from events in dense historical
narratives about international events. However, it is unwieldy and labor
intensive necessitating a search for ways to automated its application
to new corpora going forward. We present a proof of concept that ICBe
can be automated, using current generation open source large language
models, running just on consumer grade hardware. Further, our strategy
requires no fine-tuning and limited prompt-engineering, making it easy
to employ by subject experts and to modify for new topics.

This brief proof of concept is organized as follows. Section 2
formalizes the machine learning task: event abstraction from dense
historical narratives. Section 3 outlines the recent state of the art in
event coding and large language models. Section 4 describes our final
technical stack and how it optimizes across a number of constraints and
goals. Section 5 describes in detail how we map the original ICbe
ontology and human coding effort to the new ICBeLLM ontology and
corresponding prompting strategy. Section 6 evaluates the performance of
the solution, starting with precision demonstrated by a full
visualization of the entire Cuban Missile Crisis and corresponding
codings, then detailing the results of automated quality assurance
steps, moving to recall against the original human coded ICBe dataset,
and ending with a deep dive into the kinds of errors observed between
the automated and human codings. Finally, Section 7 concludes by
situating ICBeLLM in the greater context of the rapid improvement in
natural language processing in the recent past and near future.

\section{Task Definition and Domain}\label{task-definition-and-domain}

Our task is event coding from historical narratives. Event coding is an
act of abstraction, both information extraction and summarization.
History suffers from the coastline paradox, where there more finely you
measure the more detail you will necessarily find. Event coding is
therefore both a judgement about what happened and also a judgement
about at what level of detail to summarize that information. Formally, a
historical episode, H, is demarcated by a period of time,
\([T_{start}, T_{end}] ∈ T\), a set of Players \(p ∈ P\), and a set of
behaviors they undertook during that time \(b ∈ B\). International
Relations, \(IR\), is the system of regularities that govern the
strategic interactions that world actors make during a historical
episode, given their available options, preferences, beliefs, and
expectations of choices made by others. We observe neither H nor IR
directly. Rather the Historical Record, \(HR\), produces documents,
\(d ∈ D\), containing some relevant and true (as well as irrelevant and
untrue) information about behaviors that were undertaken recorded in the
form of unstructured natural language text. The task is to combine
informative priors about IR with an unstructured corpus D to produce a
series of structured discrete events, \(e ∈ E\), that have high
coverage, precision, and recall over what actually took place in
history, \(H\).

The ICBe project chooses the sentence-event as the discrete unit of
detail for a historical narrative about a large historical episode. Each
sentence can provide new information about an event, defined as a
actor-behavior pair. The ICBe project allowed for up to three distinct
events to be introduced in a sentence. We evaluate 12 nodes of the
ontology that can be used to describe an event. The ICBe ontology
recognizes three overarching classes of events: Think, Say, and Do. Do
events describe a physical action by one or more actors (Do Actor A, Do
Actor B, Do Behavior). Say events describe a communication by one or
more of the speaker actors to possibly one or more audience actors (Say
Actor A,Say Actor B, Speech Behavior). Often a say event will be about
one do event, e.g.~making a threat to invade. Think events provide
information about a cognition by one of the actors, e.g.~experienced the
start of a crisis period (Think Actor A, Thought Behavior). The details
of each do event are further recorded and we evaluate 4 pieces of
information collected (Units, Domains, Forces, Fatalities).

\section{State of the Art}\label{state-of-the-art}

Currently, the pace of development in large language models is frenetic,
with major commercial and open source releases every few months.
Evaluations are barely keeping pace, and even less so evaluations within
specific domains like political science. One such recent review
evaluates both current commercial (Claude-2, GPT-4, PALM-2) and open
source models (Llama 2, DistillRoberta) on the task of open ended survey
responses (Mellon et al. 2023). It finds high agreement between human
coders and several commercial LLMs but low performance for currently
reigning open source option Llama 2.

The most recent innovations in traditional internal event coding are
based on last generation LMs (BERT, Roberta, etc.). One is the POLECAT
dataset which is the successor to Integrated Conflict Early Warning
System (ICEWS) (Halterman, Schrodt, et al. 2023; Halterman, Bagozzi, et
al. 2023). Another is an application of a subset of the Armed Conflict
Location \& Event Data Project (ACLED) ontology to articles (Gupta and
Jamatia 2023). Likewise, within the field of natural language
processing, there is a tradition of news events based benchmarks more
broadly such as CASE 2021 event classifications (Kent and Krumbiegel
2021). Perhaps most similar to event abstraction is news summarization
(Zhang et al. 2023)

More broadly, there is a race to benchmark the capabilities (Chang et
al. 2023) and failure modes (Huang et al. 2023) of LLMS. There is work
that seeks to benchmark LLMs applied within specific domains,
e.g.~Recommendations and reviews (Liu et al. 2023), planning (Valmeekam
et al. 2023), factualness (S. Chen et al. 2023), external tool use (Li
et al. 2023), complex/ill defined tasks (Santu and Feng 2023), law (Guha
et al. 2023), pre-engineering exam knowledge (Arora, Singh, and Mausam
2023). Other work seeks to find the domain boundaries and understand
issues like model hallucination when out of distribution (Guha et al.
2023). Finally, others are concerned with the issue of contamination,
and that no easily available benchmarking text can easily be proven not
to be in the training data for an LLM prior to benchmarking (Sainz et
al. 2023).

\section{Model Selection and Prompting
Strategies}\label{model-selection-and-prompting-strategies}

Given recent advances, we have strong reason to believe that our task
should be within the reach of at least commercial LLMs. More uncertain
is whether or not current generation open source LLMs running on
consumer hardware can solve our task and thus bring international event
coding within reach for social scientists everywhere and forever (H.
Chen et al. 2023). For the same reason, we are focused on a solution
that does not require fine-tuning and instead relies on prompts that can
be changed and expanded quickly and by those with only area knowledge.
We seek a proof of concept that we have crossed the threshold of
democratized event coding.

The details of our solution are as follows. Our base model is the 70b
parameter variant of Meta's open source Llama 2 model.\footnote{See
  Roumeliotis, Tselikas, and Nasiopoulos (2023) for a survey of use
  cases across early adopters of this model.} We employ an instruction
fine-tune of it called Platypus2-70B-instruct which is better suited to
solving specific tasks than the base model which is designed primarily
for interactive chat (Lee, Hunter, and Ruiz 2023). To fit within our
compute/memory budget of two high end commercial graphics cards
(e.g.~NVIDIA 3090/4090s), we use a 4 bit precision version quantitized
by AutoGPTQ (Frantar et al. 2023). We recognize for some researchers,
particularly students, this may still be at the high end of a compute
budget but is still superior to commercial API prices (e.g.~GPT-4's API)
where this budget would buy one time processing of only about 100 pages
of text.\footnote{Currently retailing for 3-4 thousand dollars new and
  1-2 thousand dollars used. At current GPT-4 API pricing that would
  translate into a budget of of about 66,666 tokens or roughly 100 pages
  of text.} We also expect this disparity to only shrink as smaller
models improve.

In developing prompt strategies, we seek to optimize two sometimes
conflicting priorities of performance and efficiency. We need prompt
strategies that can solve the information extraction task but at a wall
clock time that's feasible for a researcher with consumer hardware.
There are fundamentally only two inputs we can optimize. The first is we
can manipulate the content and length of the input prompt. Shorter
prompts are faster to process but contain possibly less information for
solving the task. The second is we can manipulate the content and length
of the output of the model. Again shorter outputs are faster to process,
but LLMs have no scratch memory to reason with apart from the sequence
of tokens and so performance can sometimes be increased by letting the
model `reason out loud' (Kojima et al. 2022).

During the course of our prompt engineering we made the following
qualitative observations. First, requesting very long detailed outputs
is difficult even if they fit in the length of the context window. For
example, a prompt that performs sentence splitting works for paragraphs
better than a high performance NLP framework (spaCy) but fails at many
paragraphs at a time. Second, longer inputs are fine, and in fact
sometimes necessary if the context is required for name disambiguation
for example, as long as the prompt includes measures to focus and remind
the LLM of the task and most relevant parts. Third, we found that it was
far more effective to move instructions from the `question' side of the
prompt to the `answer' side, effectively pre-drafting the LLM's response
for it. Instead of instructing the LLM to ``Answer with a single
number'', it is better to begin drafting the response as ``My answer
takes the form of a single number. That number is\ldots{}'' It is more
constraining for the LLM to finish a predrafted answer than it is to
pick an answer that can either follow or ignore the question. Fourth,
multiple choice is the most efficient prompting strategy, requiring
emitting only one or a few tokens.

\section{Executing the ICBe Ontology with
LLMs}\label{executing-the-icbe-ontology-with-llms}

The defining characteristic of the ICBe ontology is that it seeks to
capture as much detail as possible along with the temporal ordering of
that detail so that a full timeline of events can be reconstructed from
a single very dense historical narrative. The narrative is broken into
paragraphs, into sentences, into events, and then if necessary into
compound events such as a speech event threatening to commit a
action-event in the future. Human coders were required for this
complicated ontology because simple dictionary based NLP methods are
unable to correctly disentangle and disambiguate that level of detail
from dense text.\footnote{Previously, most systems relied on simple
  keyword matching in news headlines.} They key question is whether LLMs
can bridge this last mile of human required steps, and if so how.

We map each of the substantive ICBe coding tasks to an LLM prompting
strategy (Table 1). The main innovation is to use the LLM's ability to
generate unstructured text to break the narrative into finer and finer
events and then code the details of those events as a final step. First,
the narrative is broken into paragraphs with simple regex and those
paragraphs are processed by the LLM into sentences. Second, if a
sentence describes more than one event, the LLM disaggregates and
composes separate complete sentences around each distinct event. This
step greatly improves on the original ICBe coding effort, where multiple
events had to be tied to entire sentences strings because narrowing it
down to specific text spans was too labor intensive. Third, each event
is checked for a thought or speech behavior which might indicate a
compound event. If one is found, the LLM further rewrites the event into
a primary and secondary event, linking the two, but allowing each to be
coded normally in the full workflow. Finally, a series of multiple
choice and open ended responses extracts information about each event in
accordance with the original ontology. After some steps we perform a
quality assurance check where we move from selecting an answer to
reviewing whether an answer is correct.

\includegraphics{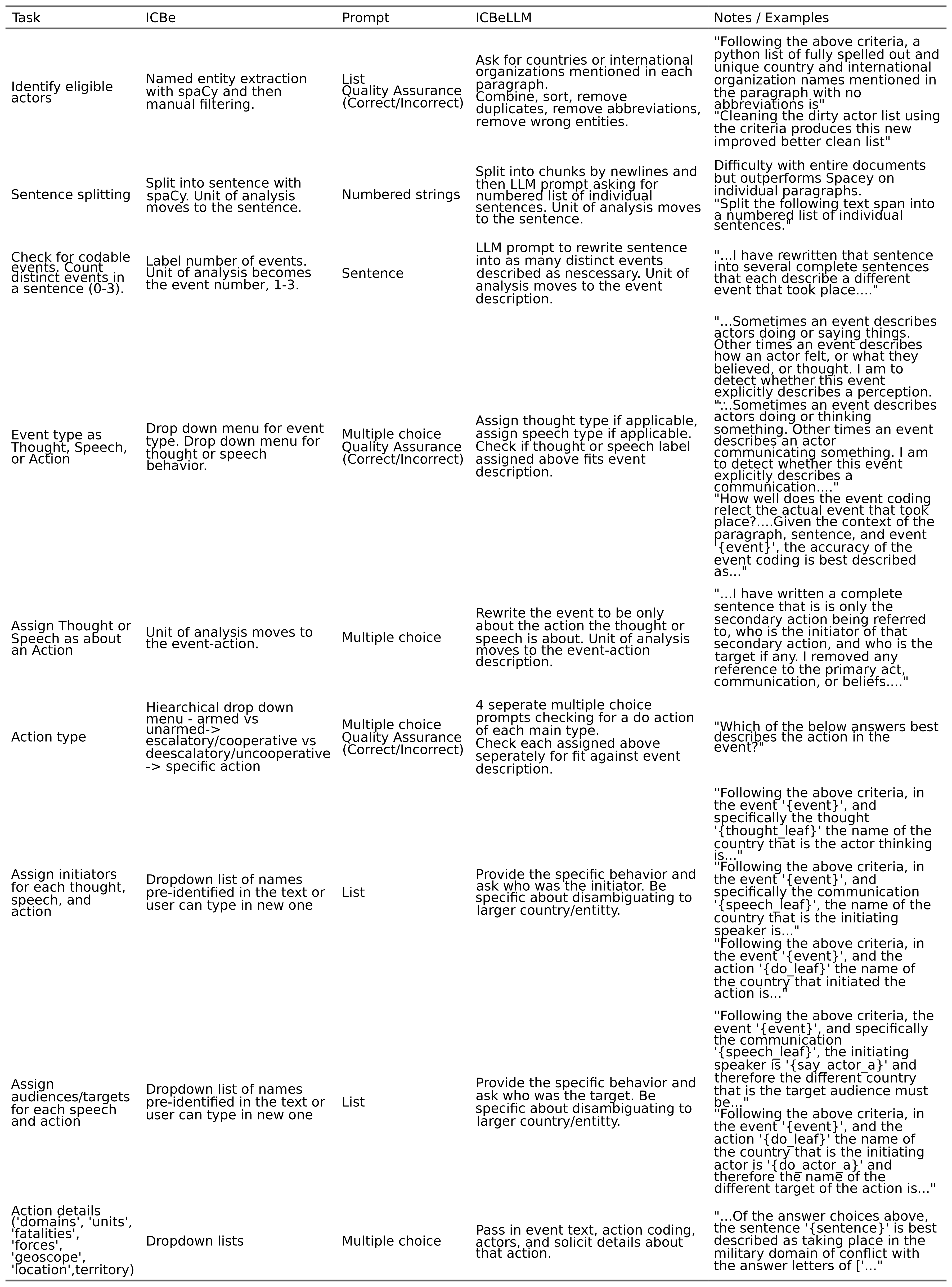}

\section{Evaluation}\label{evaluation}

Performance evaluation for an information abstraction task poses many
unique challenges because not only are we comparing the extraction of
information, we are comparing the style/level of summary chosen. Two
codings, and even two human coders, might disagree in at least 5 ways:
(1) the content of the sentence implies different events, (2) the
sentence implies more events for one than the other, (3) they agree on
the event but disagree on the level of detail supported by the text, (4)
they agree on the event but disagree on the level of summarization that
is desirable (what information to throw away), or (5) they agree on the
event but choose two different substantively equivalent ways of phrasing
it, e.g.~a threat to do something unless a concessions is provided
versus a promise not to do something if a concession is provided.

\subsection{Precision on a Qualitative Case Study - Cuban Missile Crisis
(1962)}\label{precision-on-a-qualitative-case-study---cuban-missile-crisis-1962}

We therefore begin first qualitatively with an examination of ICBeLLM's
precision relative to the original text of a crisis narrative, the Cuban
Missile Crisis. We ask to what degree the content of the source material
is accurately and completely reflected in the event codings. We utilize
a visualization that allows quick comparisons between the two. In the
table below, the first column represents the original source text of the
narrative, the second column represents the ICBeLLM's splitting and
rewriting of the sentence into individual events and compound events.
Compound events are represented as an event description immediately
followed by a second row leading with ``↳'' showing the action that the
speech or thought refers to. The third column represents the ICBeLLM's
final event coding represented with an easy to skim iconography.
Initiators are represented by their national flags, followed by a
behavior of some kind, followed by the flags of targets of that
behavior, and then for physical actions a number of icons reflect the
details of that action. For example, the domain of conflict is
represented by appropriate Mincreaft blocks (earth for ground, water for
sea, etc.), an icon of a military unit for the types of forces employed
(plastic army men for troops, a rocket for missiles, etc.), a skull and
crossbones for casualties, and a soldier silhouette for a nonzero number
of military troops involved.

The Cuban Missile Crisis took place between the United States, Cuba, and
the Soviet Union, primarily in October of 1962, over the United States's
recent attempted invasion of Cuba and the Soviet Union's deployment of
nuclear weapons to the island. We provide ICBeLLM's full coding of the
crisis below. Overall we find ICBeLLM's precision and coverage of the
Cuban Missile Crisis to be high quality. It recovers a number of tricky
events that stumped earlier NLP systems and motivated the creation of
ICBe's complicated ontology. Consider for example some of the following
key crisis points. The U.S. discovery of Soviet troops in Cuba and its
response of full mobilization (sentence 8) is correctly disaggregated by
ICBeLLM into a compound event, U.S. discovering a fact, and that fact is
the Soviet Union deployed troops to Cuba, and a second physical action
of the U.S. mobilizing it's army targeting Cuba. It further correctly
coded each of the actors, actions, and additional details of domain and
forces types. This sentence is an example where an author's desire to be
efficient and entertaining with words for a human audience creates
significant obstacles for simpler dictionary based NLP methods. In
contrast, a failure is found in the very tricky detail of the major
proposed trade (sentence 22) where the Soviet Union offered to remove
the missiles in exchange for an end to the quarantine and no invasion
pledge. Most of the actions and ancillary do details are coded correctly
but the actor directions becomes confused, and the idea of a promise to
not invade became lost.

\newpage{}

\begin{figure}[H]

{\centering \includegraphics{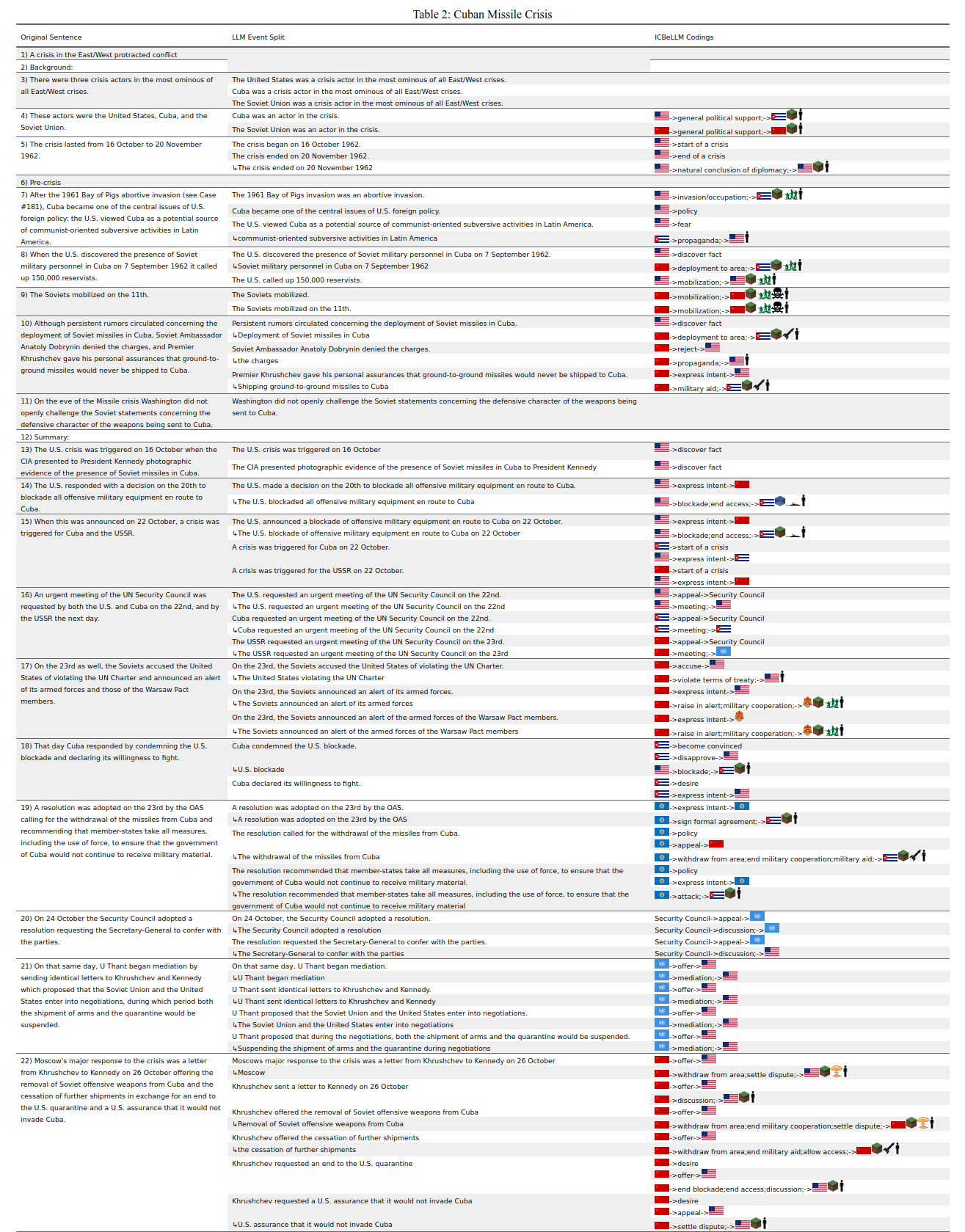}

}

\caption{Cuban Missile Crisis (Part 1)}

\end{figure}%

\newpage{}

\begin{figure}[H]

{\centering \includegraphics{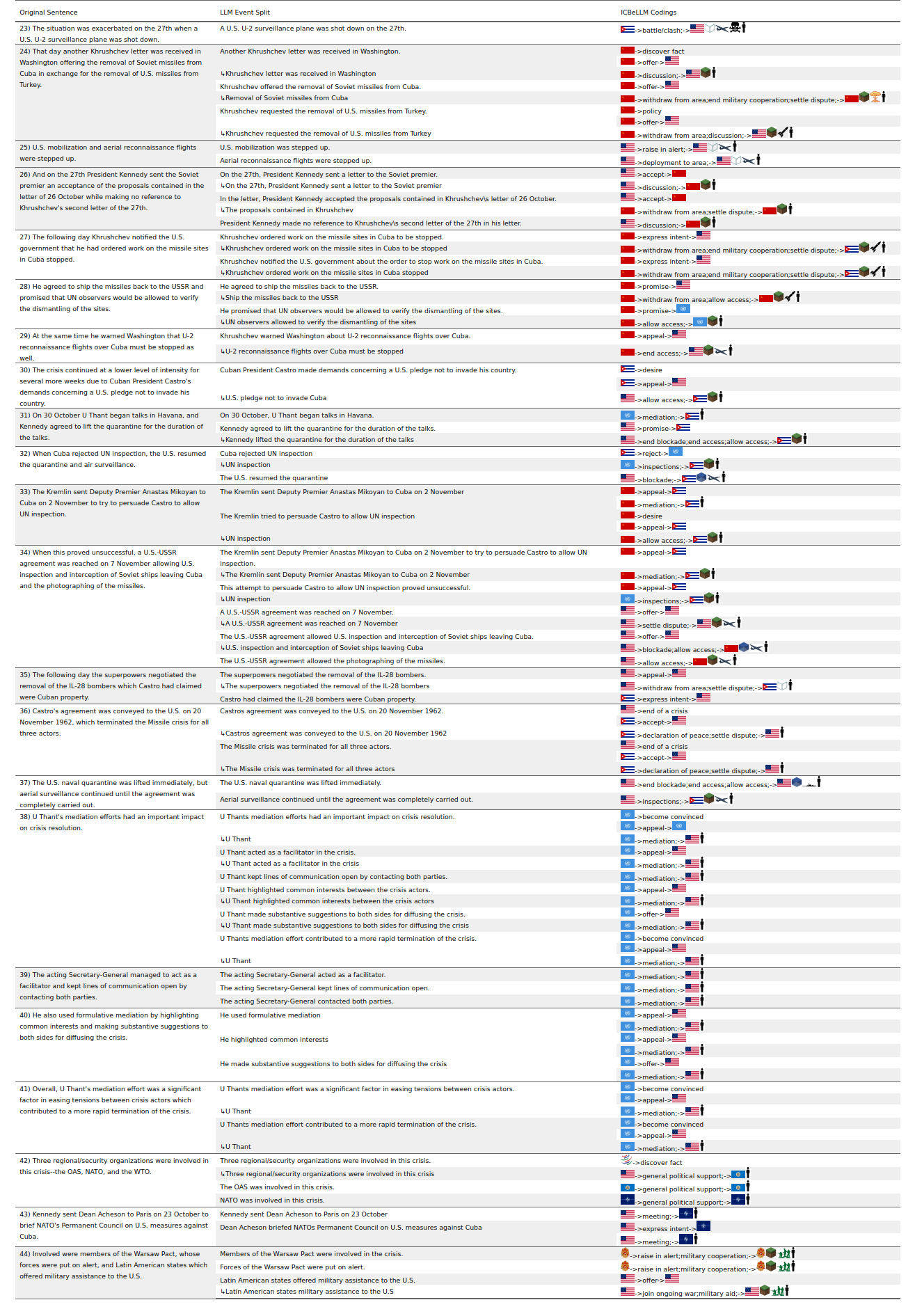}

}

\caption{Cuban Missile Crisis (Part 2)}

\end{figure}%

\subsection{Quality Control}\label{quality-control}

Our multiple choice prompting strategy is designed with speed in mind,
requiring emitting only one or a few tokens. For accuracy we would have
preferred a more elaborate prompting strategy like chain of thought,
allowing the LLM to reason `out loud' and correct earlier mistakes. As a
compromise, we employ a second round of multiple choice prompting that
asks only to evaluate the previous answer as correct or incorrect. The
QA process accepts 87\% (rejects 13\%) of all answers which large
variation indicating parts of the ontology are particularly difficult.
Almost half of all answers from the prompt asking about do actions by
unarmed actors that were uncooperative are rejected at this stage. Not
only does the QA step serve as a last line of defense against incorrect
answers, it provides a diagnostic for checking and improving prompts
without peeking at the final evaluation data. Making that comparison to
ground truth human labels is what we turn to next.

\begin{center}
\includegraphics{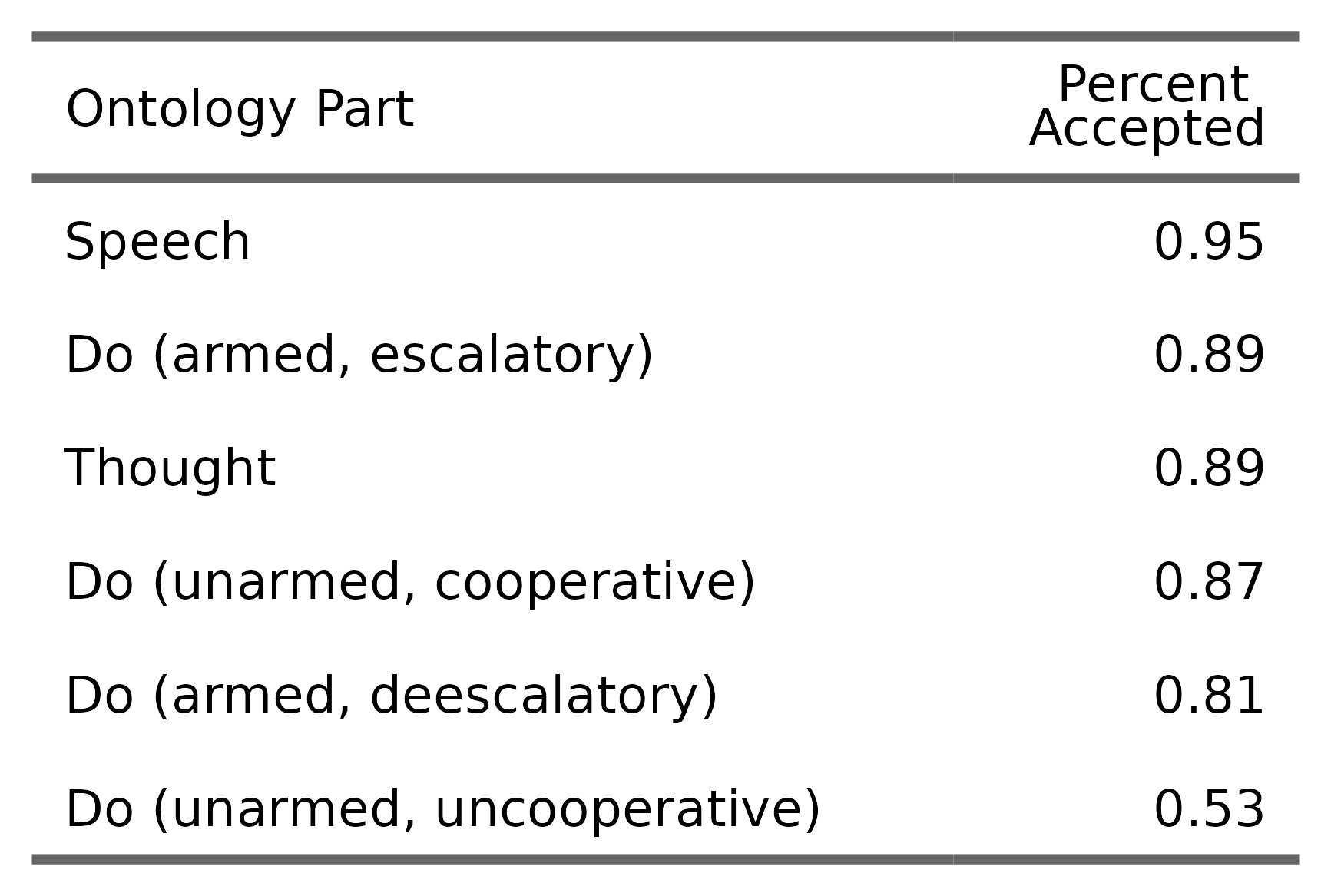}
\end{center}

\subsection{Recall}\label{recall}

While the ICBeLLM codings appear to have face validity in the
qualitative case study, we want to move toward a more quantitative
comparison of performance across the full corpus. We therefore estimate
the recall of ICBeLLM against the original ICBe human codings. We
evaluate ICBeLLM against ICBe v1.1.\footnote{Retrieved August 20, 2023,
  https://github.com/CenterForPeaceAndSecurityStudies/ICBEdataset/).}
The two do not have an exact one to one mapping because ICBeLLM is
produced at a higher level of resolution, the crisis-sentence-event and
ICBe was measured only at the crisis-sentence (with up to 3 different
events produced per sentence but no information about which part of the
sentence). Our strategy is to evaluate recall at the sentence-token unit
of observation. We define recall as the probability that a
sentence-token emitted by a human was also emmitted by the system,
\(Pr(Token_{LLM} | Token_{H})\). More concretely, we consider it a hit
if ICBeLLM emitted the same fact about a sentence, e.g.~the name of an
actor, a behavior employed, a number of casualties etc, and we consider
it a miss if ICBe emitted a fact about that sentence that was not
emitted by ICBeLLM. We further normalize and clean both sets to reduce
false misses do to unimportant differences like capitalization,
abbreviation, etc.

How to choose which of the many possible aggregations should be
considered the human ``baseline'' for this task is not trivial. The ICBe
project evaluated multiple approaches to human coding and aggregating
across disagreements and they provide important insights into the
difficult of the task and possible sources of variation.\footnote{https://github.com/CenterForPeaceAndSecurityStudies/ICBEdataset/raw/master/replication\_paper/arxiv\_draft/appendix.pdf}
Each sentence of each crisis was typically coded by at least 2 trained
coders and 2 novice coders, with a third expert coder assigned as a tie
breaker for sentences with disagreements. Intercoder agreement varied
wildly between types of coders and parts of the ontology, with experts
averaging about 65\% overall (as high as 80\% and as low as 20\% across
the ontology), and novice coders averaging only around 30\% (ranging
from 0\% to 40\% across the ontology). Intercoder agreement was
correlated with coder's own self reported confidence in their scores,
and reasons given for low confidence split about evenly between a lack
of information or confusing writing in the source text and lack of
coverage of the ontology over that specific type of detail in the text.
It was further observed that coders had individual styles, choosing
different levels of specificity when abstracting events. In short, the
task of very fine grained event abstraction is not simple even for
humans because even two trained people can look at the same narrative
and find different details to be most salient.

As with the ICBe project, we focus on areas of high agreement where
multiple coders all found the same detail salient and mapped in the same
way. We employ the ICBe ``Agreed-Wide'' version where the unit of
analysis is the crisis-sentence-actor-event type and it includes only
codings with (1) at least one expert coder's vote and (2) and a majority
of either experts or novice coders votes. We further filter down to one
event per sentence, choosing the one with the most coded information.
This leaves 126,399 tokens receiving wide agreement across 10,291
crisis-sentences. We find about 72\% recall overall which is a
comparable rate of disagreement between ICBeLLM and ICBe as between
expert human coders within ICBe and much higher than among novice
coders.

We disaggregate recall by part of the ontology in the table below.
Recall ranges from 27\% to 87\% across the ontology. We make several
broad observations. First, ICBeLLM is not yet optimized for emitting
more than one token when asked and so it will mechanically have lower
recall on questions that human coders tended to choose many tags,
e.g.~actors. All four of the top performing parts, domains, fatalities,
territory, and units tend to have only one human tag. Second, it was
easier to assign the initiator of a speech or do than it was to assign
the target. Third, questions with options that were more self-evident
performed better, such as speech behaviors like threats and offers,
where questions which were more idiosyncratic to the ICBe coding rules
performed poorly, e.g.~thought behaviors like fear, policy, discovery,
etc. We did not fine-tune, provide few shot examples, even definitions
in most cases and so this is unsurprising.

\begin{center}
\includegraphics{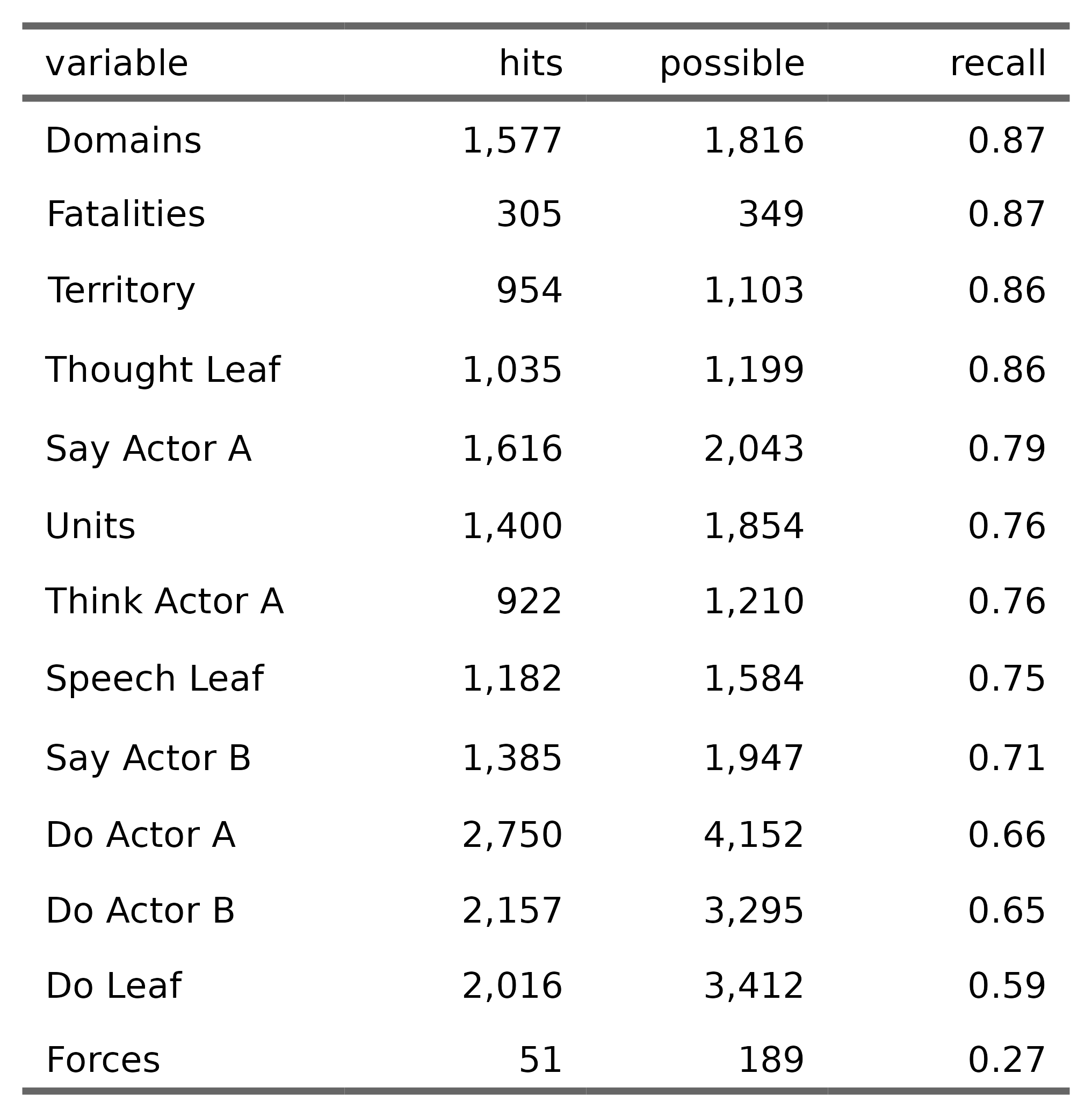}
\end{center}

\subsection{Confusion}\label{confusion}

For a relatively small subset of sentences, we can dig into the specific
disagreements between ICBeLLM and the ICBe human consensus. We only have
an unambigious 1 to 1 match between ICBe and ICBeLLM when both only
assigned a single behavior token to a sentence. That sample is both
smaller (3,228/10,573 or 30\% of all eligible for recall) and
unrepresentative because it implies simpler sentences that produced
fewer events. Still we find it useful for understanding which parts of
the ontology are performing the worst and if the confusion between
labels substantively impacts our interpretation of the final codings.

We begin first with the simplest part of the ontology, thoughts, which
only have an initiating actor and just 12 possible behaviors. We find
the largest sources of confusion to be perfectly reasonable. The two
most common types of thoughts are the start and end of a crisis, and the
start of the crisis is most frequently confused by ICBeLLM with the
three highly similar behaviors of an actor experiencing a fear, becoming
convinced of a fact, or discovering a fact. These regularly appear in
the crisis narratives as the why for a crisis beginning.

\includegraphics{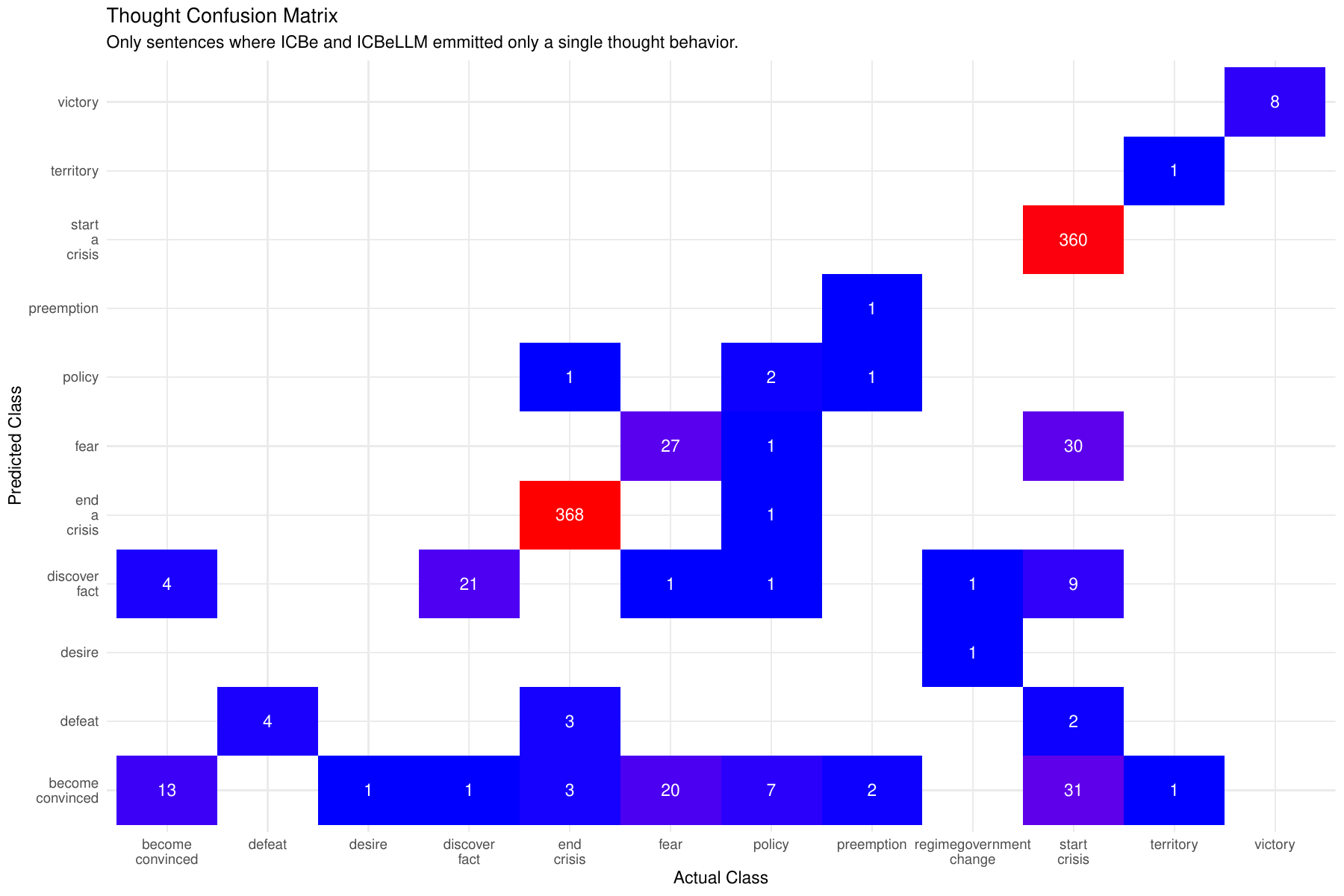}

The next most complicated behavior of speech shows greater but
understandable confusion between behaviors. The most common speech type
is appeal, which is most often confused for an offer. Disapprovals are
confused with appeals or accusations. Rejections are confused for
expressions of intent. Ultimatums are confused for threats and
expressions of intent. Offers are confused for promises. And so on. We
noted this problem in the original ICBe project that the boundary
between labels was not always well defined, and that coders developed
unique styles exploiting synonymy of the ontology.

\includegraphics{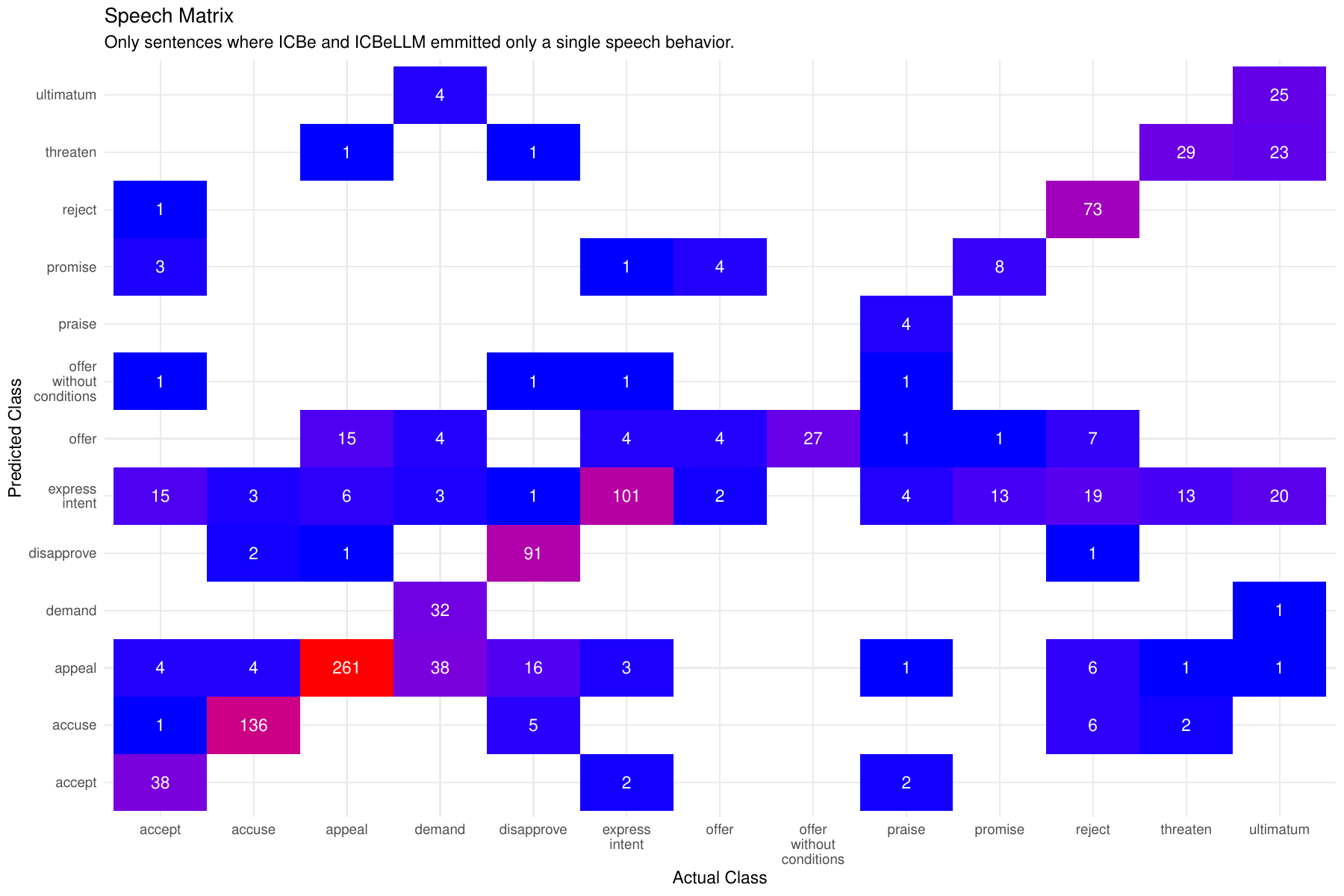}

Finally we look at confusion of the do labels, paired down to only those
that appear in ICBe 5 or more times as there are many possible do
behaviors. Again the greatest sources of confusion would not lead to
major substantive consequences. Attacks are most commonly confused with
battles/clashes, bombardments, continuation of previous fighting, etc.
Meetings are most commonly confused for discussions and vice versa.
Deployments to an area are confused for invasions/occupations. In short
the kinds of errors we see are what we would hope to see if our overall
approach was correct but our ontology and definitions need to be
clearer.

\includegraphics{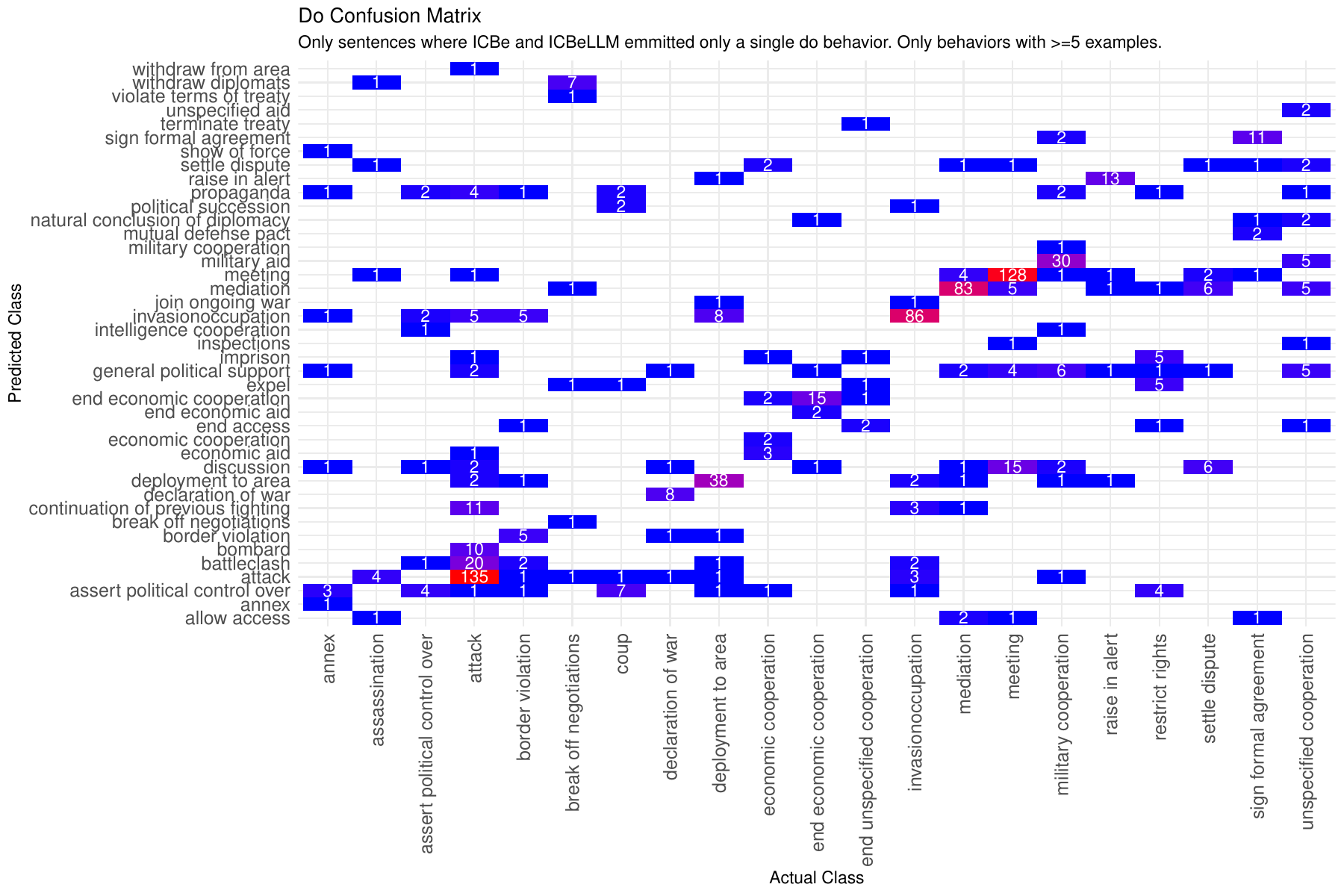}

\section{Conclusion}\label{conclusion}

We investigated whether the current generation of open source LLMs are
able to take over previously human only international event coding and
extraction tasks. We attempted to automate a very complicated and high
quality ontology of international events which required human labeling
to perform and could not be achieved via the last generation of
dictionary based NLP methods. We evaluated the precision of the
automated codings on a detailed qualitative case study and the recall
over the full corpus of tens of thousands of events across nearly five
hundred historical crises. We found that we have entered the age of at
home event coding, where researchers and students will be able to mine
text for advanced concepts and regularities nearly for free and forever
with open source consumer hardware.

We stress that this was only a successful proof of concept, and just the
floor for what is possible. We concluded our experiments at an
interesting inflection point in open source LLM development. Very large
open source models capable of competing with commercial alternatives had
just become both sufficiently capable and compressible to run at all on
consumer level hardware. This home setup works but is not fast and many
of our prompt strategies are geared around minimizing the number of
tokens emitted and multistage feedback and interaction with the LLM.
Already another generation of open source models are arriving which
rival commercial LLMs but at a fraction of the memory and compute
requirements. This will enable more sophisticated prompt strategies
which allow for reflection, interrogation, and brainstorming and will
greatly improve both the recall and precision for this kind of event
extraction task. We, for example, employed very simple post-coding
quality assurance checks. These and many others will become commonplace
as the time and energy cost per token continues to fall year after year.

The ICBe project is an example of right place at the right time. It
built an ontology with very high coverage and detail at the risk of
being too unwieldy to justify another large commitment in human coders
on new documents. But it completed just in time for the appearance of
open source large language models that can easily implement the coding
at scale. Both the research investment in ontology design and human
labeling of examples made translation zero shot prompts for an LLM a
relatively simple exercise.

Given the pace of technological advancement, we expect event projects to
continue this trend of shifting human labor towards ontology design.
Undervalued definition and codebook authoring has been rebranded as
`prompt engineering' for a new generation of systems with a LLM in the
loop rather than a hired human coder. No matter the justification, this
is a positive development for empiricism in social science. Definitions
can now be chosen at the time of the analysis and rerun overnight rather
than pre-committed to years earlier in the cycle of grant raising,
hiring, training, project management, and eventually publication. This
reduction in the lag between theory generation and data collection from
an unexpected source is a critical and welcome development in the
maturation of social science on international events.

\section*{Works Cited}\label{works-cited}
\addcontentsline{toc}{section}{Works Cited}

\phantomsection\label{refs}

\end{document}